\documentclass[fleqn,10pt]{wlscirep}

\usepackage{bm}
\usepackage{graphicx}
\usepackage[squaren]{SIunits} 
\usepackage{hyphenat}
\usepackage{hyperref}
\usepackage{color}
\usepackage[capitalise]{cleveref}
\crefname{subsection}{subsection}{subsections}
\usepackage{multirow}
\usepackage{rotating}
\usepackage{makecell}

\definecolor{pink}{RGB}{255,0,255}

\definecolor{ss_color}{rgb}{1,0,0}

\definecolor{yz_color}{rgb}{0,0,1}

\definecolor{cl_color}{rgb}{1,0,1}

\definecolor{new_color}{rgb}{0,0,1}

\title{Bright-light detector control emulates the local bounds of Bell-type inequalities}

\author[1,2,3,*]{Shihan Sajeed}
\author[1,3]{Nigar Sultana}
\author[4,5]{Charles Ci Wen Lim}
\author[6,7,8,3]{Vadim~Makarov}

\affil [1] {Institute for Quantum Computing, University of Waterloo, Waterloo, ON, N2L~3G1 Canada}
\affil [2] {Department of Physics and Astronomy, University of Waterloo, Waterloo, ON, N2L~3G1 Canada}
\affil [3] {Department of Electrical and Computer Engineering, University of Toronto, M5S~3G4, Canada}
\affil [4] {Centre for Quantum Technologies, National University of Singapore, Singapore}
\affil [5] {Department of Electrical \& Computer Engineering, National University of Singapore, Singapore}
\affil [6] {Russian Quantum Center, Skolkovo, Moscow 121205, Russia}
\affil [7] {Shanghai Branch, National Laboratory for Physical Sciences at Microscale and CAS Center for Excellence in Quantum Information, University of Science and Technology of China, Shanghai 201315, People's Republic of China}
\affil [8] {NTI Center for Quantum Communications, National University of Science and Technology MISiS, Moscow 119049, Russia}
\affil[*] {shihan.sajeed@gmail.com}

\date{\today}
\begin{abstract}
It is well-known that no local model---in theory---can simulate the outcome statistics of a Bell-type experiment as long as the detection efficiency is higher than a threshold value. For the Clauser-Horne-Shimony-Holt (CHSH) Bell inequality this theoretical threshold value is $\eta_{\text{T}} = 2 (\sqrt{2}-1) \approx 0.8284$. On the other hand, Phys.\ Rev.\ Lett.\ 107, 170404 (2011) outlined an explicit practical model that can fake the CHSH inequality for a detection efficiency of up to $0.5$. In this work, we close this gap. More specifically, we propose a method to emulate a Bell inequality at the threshold detection efficiency using existing optical detector control techniques. For a Clauser-Horne-Shimony-Holt inequality, it emulates the CHSH violation predicted by quantum mechanics up to $\eta_{\text{T}}$. For the Garg-Mermin inequality---re-calibrated by incorporating non-detection events---our method emulates its exact local bound at any efficiency above the threshold. This confirms that attacks on secure quantum communication protocols based on Bell violation is a real threat if the detection efficiency loophole is not closed.
\end{abstract}

\begin{document}
\flushbottom
\maketitle

\thispagestyle{fancy}

\section*{Introduction}

More than $50$ years ago, John Stewart Bell showed that any physical theory based on the assumptions of locality (i.e.,\ nothing can communicate faster than light) and realism (i.e.,\ physical properties of an object are fixed and pre-defined) must satisfy a set of statistical criteria called Bell inequalities \cite{bell1964}. That is, if a Bell-type experiment is performed and the results show a violation of a Bell inequality, then the underlying physical process cannot be explained by a local theory.  This kind of tests are called Bell tests and the violation of the inequality is called Bell violation. Since the earlier demonstrations utilizing cascade decays in atoms~\cite{freedman1972,aspect1981,aspect1982,aspect1982a}, Bell violations have been observed in tests utilizing nonlinear optical processes \cite{weihs1998,giustina2013,christensen2013,poh2015}, ions \cite{rowe2001}, neutral atoms \cite{hofmann2012}, Josephson junction \cite{ansmann2009} and solid state qubits \cite{pfaff2012}. The implications of the Bell test not only change our understanding of nature, but also find application in device independent (DI) quantum communications \cite{ekert1991,mayers1998,barrett2005}, randomness generation and amplification \cite{colbeck2006,pironio2010,colbeck2012}, DI-verified quantum computation \cite{gheorghiu2015,hajdusek2015}, certifying quantum devices \cite{mayers2004,mckague2010,pironio2010} and DI bit commitment \cite{aharon2016}. Entanglement, a necessary precondition for unconditional security \cite{lo1999,curty2004a} in quantum key distribution, can also be certified from the violation of a Bell inequality, independently of the underlying implementation details. This paves the way for the device-independent tests of security \cite{acin2007,acin2006}. However, for the observed Bell violation to be conclusive, it is important that the Bell test is loophole-free. 

More specifically, a loophole-free Bell test is an entanglement experiment that requires multiple implementation loopholes such as the detection, locality, and measurement-independent loopholes to be closed simultaneously. Here, we focus on the detection loophole, and defer the rest to Ref.~\cite{brunner2014}. In general, the detection loophole is a scenario in which the observed Bell violation (a test statistics) is no longer reliable as the measurement sample and may not be a true representative of the population (i.e., the entire measurement statistics). Crucially, this situation commonly happens in practice as practical detectors have finite detection efficiencies and hence one could end up with samples that are non-representative. While the detection loophole is not an issue for non-adversarial settings, the same is not true for the case of quantum cryptography since an adversary can take advantage of it to come up with a local model to \emph{fake} Bell violations~\cite{gerhardt2011a}. For this reason, much effort has been devoted to closing the detection loophole in practice.

How a local model can theoretically simulate non-local correlations---taking advantage of the detection loophole---has already been reported in the literature~\cite{gisin1999,larsson1999}. However, methods of experimentally implementing such correlations using practical means have been rarely discussed, despite its importance in practical quantum cryptography. The state-of-the art method is arguably that of Ref.~\cite{gerhardt2011a}, where the authors demonstrated how an adversary could implement a local model using existing optical detector control methods to violate a Bell inequality for active basis choice schemes. However, their local model is effective only for a detector efficiency of up to $\eta = 0.5$, while theoretically it is possible to fake the inequality for a threshold efficiency of up to $\eta_{\text{T}} = 2 (\sqrt{2}-1) \approx 0.8284$ (here, efficiency $\eta$ refers to the probability that one party observes a conclusive outcome given a measurement is made). In this article, we discuss how to experimentally close this gap and fake the violation at higher efficiencies. More specifically, we show how existing optical detector control methods~\cite{lydersen2010a,gerhardt2011,liu2014} can be exploited to both fake the violation of the standard Clauser-Horne-Shimony-Holt (CHSH) Bell test all the way up to its threshold efficiency and simulate the local bound of the more general Garg-Mermin Bell test. Our results point out once again that when Bell tests are performed for certifying randomness, guaranteeing security in quantum communications, or detecting non-locality, they should either be performed with an efficiency at which the test is robust against detection loopholes, or should use the bound given by more general inequalities (for example, \cref{eqn2} below). Otherwise, existing optical detection control methods may allow to implement a local model to simulate the results of the test.

The article is organized as follows. First we outline the assumptions and methodology of the Bell test that we consider in this article. Then we present several local models that allow an adversary to implement a practical setup to fake the Bell test or emulate the local bounds given by the inequalities. Then we make our conclusion.

\section*{Assumptions for Bell test}
\label{assumptions}

\begin{figure}
	\centering
	\includegraphics[width=0.5\columnwidth]{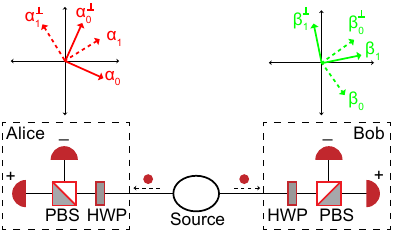}
	\caption{Setup for a CHSH test. The measurement angles shown are arbitrary. Here, for each party, the set of possible outcomes is given by $\{+,-,?\}$, where ``$+$'' (``$-$'') are assigned when only the lower (upper) detector clicks and the other detector is silent. ``$?$'' is assigned when none of the detectors registers a detection.}
	\label{fig:setup}
\end{figure}

The experimental setup of the CHSH Bell test for two parties with binary inputs and outputs \cite{clauser1969} is shown in \cref{fig:setup}. The test assumes that a source of entangled photon pairs sends each member of the pairs to two legitimate parties, Alice and Bob. Alice randomly measures the polarizations along direction $\alpha_0$ or $\alpha_1$ and Bob randomly along $\beta_0$ or $\beta_1$ as shown in \cref{fig:setup}. The measurement along a particular direction is performed with the help of a rotatable half wave plate (HWP) followed by a polarization beamsplitter (PBS) and two single photon detectors. This type of analyzer is called an active basis choice analyzer. The possible polarization measurement outcomes expected at Alice and Bob are $P_A \in \{\alpha_0, \alpha_0^\perp, \alpha_1, \alpha_1^\perp\}$ and $P_B \in \{\beta_0, \beta_0^\perp, \beta_1, \beta_1^\perp\}$, and they are mapped into outcomes $\{+,-,?\}\times\{+,-,?\}$; see \cref{fig:setup} for outcome assignments.

We assume that Alice and Bob are situated far apart, so that the locality loophole is closed. However, due to the finite efficiency of the detectors and optical losses in the setup, it is not possible to measure the polarization of all the photons. So, the final statistics are calculated using only the detected events, i.e.,\ events in which photons have been detected on both sides. In this case, for each pair of measurement settings $\{\alpha_i,\beta_j\}$ with $ij \in \{00,01,10,11\}$ chosen by Alice and Bob, the correlation function $E(\alpha,\beta)$ is given by
\begin{eqnarray}
\label{expect}
 E(\alpha,\beta) =  \frac{N_{\alpha,\beta}(++) + N_{\alpha,\beta}(--) - N_{\alpha,\beta}(+-) - N_{\alpha,\beta}(-+)} {N_{\alpha,\beta}(++) + N_{\alpha,\beta}(--) + N_{\alpha,\beta}(+-) + N_{\alpha,\beta}(-+)},
\end{eqnarray} 
where $N_{\alpha,\beta}(i,j)$ represents the number of coincidences with successfully detected outcome $\{i,j\} \in \{++,+-,-+,--\}$ for a particular setting $(\alpha,\beta)$. The associated CHSH Bell inequality is then
\begin{equation}\label{eqn1}
\begin{aligned}
S_{\rm{CHSH}}=  
E(\alpha_0, \beta_0) + E(\alpha_1,\beta_0) + E(\alpha_1,\beta_1) - E(\alpha_0,\beta_1)\leq 2.
\end{aligned} 
\end{equation}

Quantum mechanics predicts a maximum violation of $S = 2 \sqrt{2}$ for the setting choice $\alpha_0 = -78.75\degree$, $\alpha_1 = 56.25\degree$, $\beta_0 = 11.25\degree$, $\beta_1 = -33.75\degree$ \cite{eberhard1993}, and even stronger correlations are algebraically possible in theory leading to $S \le 4$ \cite{popescu1994}. However, as long as the efficiency of a measurement is $\eta =1$, all local models must necessarily satisfy Eq.~(\ref{eqn1}). Unfortunately, this is not true for $\eta < 1$. In particular, when $\eta$ is less than some threshold $\eta_{\text{T}}$, it is possible to devise local models that violate \cref{eqn1}. For the CHSH test described here, $\eta_{\text{T}} = 2(\sqrt{2} -1)\approx 82.84\%$~\cite{garg87}. In order to avoid this, these tests are performed in the region $\eta > \eta_{\text{T}}$. Note that the CHSH test is not the most robust Bell test as one can further reduce the detection threshold by looking at marginal correlations (or singles statistics). This is given by the Eberhard Bell inequality \cite{eberhard1993}, which has a detection threshold of $\eta_T = 2/3\approx 66.67\%$. Alternatively, one can include the `efficiency' in the inequality and recalibrate it as a function of $\eta$ as proposed by Garg and Mermin \cite{garg87}
\begin{equation}
\begin{aligned}
\label{eqn2}
S'(\eta) &= E(\alpha_0,\beta_0) + E(\alpha_1,\beta_0) + E(\alpha_1,\beta_1) - E(\alpha_0,\beta_1)\\
& \leq  \frac{4}{\eta} - 2.
\end{aligned}
\end{equation}
The recalibrated CHSH Bell inequality gives the local bound of $S'$ as a function of $\eta$, i.e., how much violation is required to certify non-locality for a given efficiency. This is shown by the solid (red) curve in \cref{fig:result_ce}. Note that, when $\eta=1$, \cref{eqn2} becomes \cref{eqn1} since the post-selected correlation set becomes the entire measurement set. Also, when $\eta \leq 2/3$, one can always set the bound to be $4$, which is the maximum value attainable by the sum of four correlation functions. Thus, a local model that can simulate \cref{eqn2} for efficiency range $ 2/3 \le \eta \le 1$ would be the optimum model to exploit detection loopholes in a Bell test. We present it in the next section.

\section*{Faking Bell inequality with improved efficiency}
\label{faking}

\begin{figure}
	\centering
	\includegraphics[width=0.57\columnwidth]{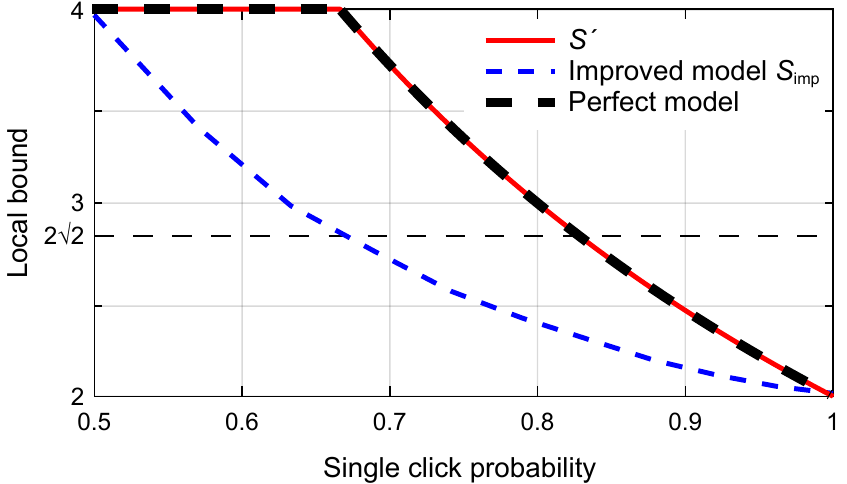}
	\caption{Local bounds for recalibrated inequality $S'$ [\cref{eqn2}], improved faking model [\cref{S_vs_n}], and perfect faking model. The quantum mechanical bound $2\sqrt{2}$ is also shown. The improved faking model achieves this bound at $\eta \approx 0.6678$ and the perfect model at $\eta = 2 (\sqrt{2}-1)$. The perfect model can fully emulate \cref{eqn2} for efficiency range $ 2/3 \le \eta \le 1$.}
	\label{fig:result_ce}
\end{figure}

For ease of understanding, we will go step by step. First, we review an existing local model that can fake \cref{eqn1} for $\eta \le 1/2$ \cite{gerhardt2011a} and point out its limitations. Then we propose a modification to this model that enables it to fake \cref{eqn1} up to $\eta \le 2/3$. We then present our perfect model that can not only fake \cref{eqn1} for $\eta \le 2 (\sqrt{2}-1)$ but also emulate the local bounds given by \cref{eqn2}. Since all three models exploit an existing detector control method---bright-light detector control \cite{lydersen2010a,gerhardt2011,liu2014}---we first recap it. 

\textbf{Bright-light detector control:} Single-photon detectors used in a Bell test may become insensitive to single photons when exposed to bright light \cite{lydersen2010a,lydersen2011c}. Even in this mode, they can produce a detection event (`click') when additionally exposed to a light pulse of intensity $I$ equal to or higher than a threshold level $I_\text{th}$. This allows an adversary Eve to have control over the detectors by tailoring $I$. For example, if the measurement basis matches that of the incoming light pulse, then all of it is incident on a single detector with intensity $I \ge I_\text{th}$ and results in a detection event. However, in case of basis mismatch, the incoming light is split between two detectors with intensity $I/2 < I_\text{th}$ (assuming a conjugate basis) and none of the detectors click. This is how the adversary can have control over detection outcomes. The feasibility of bright-light control has been confirmed numerous times, with both detectors based on avalanche photodiodes \cite{lydersen2010a,gerhardt2011,lydersen2010b,wiechers2011,sauge2011,jogenfors2015,huang2016,chistiakov2019,gras2020} and superconducting nanowires \cite{lydersen2011c,tanner2014,elezov2019}. Next, we show how an adversary can exploit it to implement a local-realistic model.

\textbf{Conditions for violation:} Let us assume that an arbitrary value of $ |S| \le 4$ needs to be simulated by the local model. Assuming symmetry for each setting combination $(\alpha,\beta)$,  this implies $|E| = S/4$. Assuming $N_{\alpha,\beta}(++) = N_{\alpha,\beta}(--) = N_\text{sim}$ and $N_{\alpha,\beta}(+-) = N_{\alpha,\beta}(-+) = N_\text{dif}$, (where $2 N_\text{sim} + 2 N_\text{dif} = 1$), Eq.~\ref{expect} can be written as
\begin{equation}
\frac{N_\text{sim}}{N_\text{dif}} = \frac{1 + E}{1 - E}.
\label{xy_rel}
\end{equation}
This implies that under the assumptions specified above, an arbitrary correlation value $E$ requires the ratio of similar to different outcomes to follow \cref{xy_rel}. For example, the quantum mechanical prediction of $S = 2 \sqrt{2}$, which corresponds to $E= \pm 1/{\sqrt{2}}$, requires 
\begin{equation}
N_\text{sim} = (3 \pm 2\sqrt{2}) N_\text{dif}.
\end{equation}
Below we describe several techniques by which an active attacker can satisfy this condition. 

\begin{table}
	\centering
	\caption{Probability of each polarization combination generated by the source in the existing faking model~\cite{gerhardt2011a}. They are normalized to maintain $2 N_\text{sim} + 2 N_\text{dif} = 1$.}
	\label{tab:first-model-polarization-probabilities}
	\renewcommand{\arraystretch}{1.5}
	\setlength{\tabcolsep}{7 pt}
	\begin{tabular}[t]{||c|c|cccc}
		\cline{3-6}
		\multicolumn{2}{c}{} & \multicolumn{4}{c}{} \\[-1.5em]
		\cline{3-6} 
		\multicolumn{2}{c}{} & \multicolumn{4}{c}{Towards Bob}\\
		\cline{3-6} 
		\multicolumn{2}{c}{} & $\beta_0$ & $\beta_0^\perp$ & $\beta_1$ &$\beta_1^\perp$  \\ 
		\cline{3-6} 
		\multirow{3}{*}{\begin{turn}{90} Towards Alice~~~ \end{turn}}
		& $\alpha_0$       & $N_\text{sim}/4$ & $ N_\text{dif}/4 $ & $N_\text{dif}/4$ & $N_\text{sim}/4$ \\ 
		& $\alpha_0^\perp$ & $N_\text{dif}/4$ & $ N_\text{sim}/4 $ & $N_\text{sim}/4$ & $N_\text{dif}/4$ \\ 
		& $\alpha_1$       & $N_\text{sim}/4$ & $ N_\text{dif}/4 $ & $N_\text{sim}/4$ & $N_\text{dif}/4$ \\ 
		& $\alpha_1^\perp$ & $N_\text{dif}/4$ & $ N_\text{sim}/4 $ & $N_\text{dif}/4$ & $N_\text{sim}/4$ \\ 
	\end{tabular} 
\end{table}

\textbf{Existing model:} A straightforward approach to force the outcomes to follow \cref{xy_rel} is to generate polarization combinations at the source with desired statistics and then force deterministic outcomes during the measurement, as done in Ref.~\citen{gerhardt2011a}. We assume each polarization combination is generated according to the probabilities given in~\cref{tab:first-model-polarization-probabilities}, where $N_\text{sim}$ and $N_\text{dif}$ obey \cref{xy_rel}. We assume the intensity is tailored to bring the bright-light control method into play, i.e.,\ matched (mismatched) bases lead to deterministic outcome with unity probability (no detection). Let's consider the case when the source generates polarization combination $\alpha_0 \beta_0$ $(\alpha_0 \beta_1)$ with probability $N_\text{sim}/4$ $(N_\text{dif}/4)$. They result in coincidences only for the setting $\alpha_0 \beta_0$ $(\alpha_0 \beta_1)$ and lead to deterministic similar (similar) outcomes with unity probability. For the remaining three setting choices, no coincidence happens and the outcomes have no effect on the correlation. This is true for all the polarization combinations in \cref{tab:first-model-polarization-probabilities}. In this way, it is possible to generate outcomes to match \cref{xy_rel} for any desired value of $E$ and achieve any value of $S$ up to $S_1 = 4$. A problem with this method, however, is that half of the time the measurement basis does not match the preparation basis and results in no detection. Thus the efficiency at each side $\eta_1 = 0.5$. This is a limitation in Ref.~\citen{gerhardt2011a}. Next, we outline how to implement an improved local realistic model with a higher detection efficiency.

\textbf{Improvement to existing model:} Above we have recapped the existing first method that leads to CHSH parameter $S_1=4$ with an efficiency $\eta_1=0.5$. We now generate a second method that leads to CHSH parameter $S_2 = 2$ with efficiency $\eta_2 = 1$. For this, let's assume that the source always sends polarization $\alpha~(\beta)$ to Alice (Bob), where $\alpha$ ($\beta$) is polarized at an angle that is midway between $\alpha_0$ and $\alpha_1$ ($\beta_0$ and  $\beta_1$). In this case, irrespective of the measurement settings, the input intensity $I$ is split at a ratio of $\cos^{2}(\phi_A): \sin^2(\phi_A) $ between the two detectors in Alice and at $\cos^{2}(\phi_B): \sin^2(\phi_B)$ in Bob. Here, $\phi_A = |\alpha_1 - \alpha_0|/2$ and $\phi_B = |\beta_1 - \beta_0|/2$. Tailoring the intensity to satisfy $I \cos^{2}(\phi) \ge I_\text{th} $ and $I \sin^{2}(\phi) < I_\text{th}$ at the respective sides ensures that only one of the detectors clicks (with outcome $+$), irrespective of the basis choice, and efficiency stays~1. This will result in $E=+1$ for each measurement setting and lead to a CHSH parameter $S_2=2$ with an efficiency $\eta_2 =1$. Note that this method (presented here for its ease of explanation) results in only $++$ outcomes. It can be symmetricized to produce all four outcomes $++$, $+-$, $-+$, $--$, which we omit for brevity. 

Thus, we have outlined two independent approaches to control $S$: the first one leads to $S_1 = 4$ with an efficiency $\eta_1 = 0.5$, while the second one leads to $S_2 = 2$ with efficiency $\eta_2 = 1$.  An adversary can then use a probabilistic mixture of these two approaches to increase the faking efficiency of the Bell test. With probability $p_1~(p_2 = 1 - p_1)$ she uses the first (second) method. The input intensity needs to be tailored to $2 I_\text{th} >  I \ge  I_\text{th}/\cos^2(\phi)$ to ensure that the first (second) method leads to detection efficiency of $\eta_1 = 0.5$ ($\eta_2 = 1$) and results in $S_1 = 4$ ($S_2 = 2$). The resultant efficiency as seen by Alice and Bob will be $\eta = \sqrt{p_1 \eta_1^2 + p_2 \eta_2^2}$ and the improved CHSH parameter will be
\begin{equation}
S_{\text{imp}} = \frac{p_1 S_1 \eta_1^2 + p_2 S_2 \eta_2^2}{\eta^2}.
\label{S_vs_n}
\end{equation}
The variation of $S_{\text{imp}}$ with $\eta$ is shown  in \cref{fig:result_ce}. The left-most point $(\eta,S_{\text{imp}}) = (0.5,4)$ corresponds to the first method with $p_2 =0$. As $p_2$ is increased, $S_{\text{imp}}$ becomes smaller with increasing efficiency and eventually becomes $(\eta,S_{\text{imp}}) = (1,2)$ at the rightmost point with $p_2 =1$. Quantum mechanical prediction $S =2 \sqrt{2} $ is obtained at $p_2 \approx 0.2612$ and the corresponding efficiency is $\eta \approx 0.6678$. This is still lower than the threshold efficiency limit $\eta_T = 2 (\sqrt{2}-1) \approx 0.8284$ for CHSH inequality. To achieve higher local bounds, one more degree of freedom needs to be introduced, as discussed in our next model.

\begin{table}
	\centering
	\caption{Possible outcomes and the corresponding probabilities for different measurement settings in the perfect model. Outcome $i j \in \{+,-,?\}\times\{+,-,?\}$ represents $i$ at Alice and $j$ at Bob. It can be verified that the conditional probability distributions are no-signalling \cite{branciard2011}.}
	\label{tab:third-model-outcomes}
	\renewcommand{\arraystretch}{1.16}
	\begin{tabular}[t]{@{\extracolsep{3ex}}c@{}c@{}c@{}c@{}c@{}c@{}}
		\hline\hline
		\multirow{2}{*}{\raisebox{0mm}[0mm][0mm]{\makecell{Polarization\\ emitted\\ from source}}} & \multirow{2}{*}{\makecell{Measurement\\ outcome}} & \multicolumn{4}{c}{\makecell{Joint probability\\ at measurement setting}} \\
		\cline{3-6}
		& & $\alpha_0 \beta_0$ & $\alpha_1 \beta_0$ & $\alpha_0 \beta_1$ & $\alpha_1 \beta_1$ \\
		\hline
		$\alpha_0 \beta_0$	& $++$	& $a$		& $b/2$	& $0$		& $0$		\\
												& $+-$	& $0$		& $0$		& $a$		& $b/2$ \\
												& $-+$	& $0$		& $b/2$	& $0$		& $0$		\\
												& $--$	& $0$		& $0$		& $0$		& $b/2$	\\
												& $?+$	& $1-a$	& $1-b$	& $0$		&	$0$		\\
												& $?-$	& $0$		& $0$		& $1-a$	&	$1-b$	\\
		\hline   
		$\alpha_1 \beta_1$	& $++$	& $b/2$	& $a$		& $b/2$	& $a$		\\
												& $+-$	& $0$		& $0$		& $0$		& $0$		\\
												& $-+$	& $b/2$	& $0$		& $b/2$	& $0$		\\
												& $--$	& $0$		& $0$		& $0$		& $0$		\\
												& $?+$	& $1-b$	& $1-a$ & $1-b$	& $1-a$	\\
												& $?-$	& $0$		& $0$		& $0$		&	$0$		\\
		\hline\hline
	\end{tabular}
\end{table}

\textbf{Perfect local model:} Now we present a perfect local model that can not only fake a violation of inequality (\ref{eqn1}) for $\eta \le 2 (\sqrt{2}-1)$ but also emulate the local bounds given by \cref{eqn2} for $2/3 \le \eta \le 1$. For this model, we make three assumptions:
\begin{enumerate}
	\item The adversary at the source always generates one of the two polarization combinations $\alpha_0 \beta_0$ and $\alpha_1 \beta_1$ with equal probability of $1/2$ each. 
	\item The adversary tailors the light intensity towards Bob in such a way that they result in a deterministic outcome with unity probability. For the ease of analysis we will assume that at Bob, the polarization $\beta_0$ ($\beta_1$) leads---with unity efficiency---to deterministic outcome ``$+$'' (``$+$'') when measured along $\beta_0$ and ``$-$'' (``$+$'') when measured along $\beta_1$ (however, any other outcomes will also do as long as they are deterministic and have unity probability). 
	\item At Alice, whenever the measurement basis matches (does not match) that of the incoming light, a deterministic ``$+$'' (random) outcome is produced with probability $a$ ($b$), and no-detection outcome with probability $1-a$ ($1-b$).  
\end{enumerate}

For details on how the above assumptions can be met in practice, please see the Methods. For each setting, the possible outcomes at Alice and Bob and the corresponding coincidence probabilities are shown in \cref{tab:third-model-outcomes}. For any measurement setting $\{\alpha,\beta\}$, the correlation function $E$ is related to $a$ and $b$ as
\begin{equation}
|E| = \frac{\frac{a}{2} +  \frac{b}{4} -  \frac{b}{4}}{ \frac{a}{2} + 2 \frac{b}{4}} =   \frac{a}{a+b},
\label{rel_eab}
\end{equation}
and the coincidence probability is
\begin{equation}
 \frac{a}{2} + \frac{b}{2} = \eta^2.
\label{rel_ab2}
\end{equation}
Solving \cref{rel_eab,rel_ab2}, we get,
\begin{equation}
\begin{aligned}
a =& 2  E \eta^2\\
b =& 2 (1-E) \eta^2
\label{ab}
\end{aligned}
\end{equation}
Thus, to emulate the local bounds in an actual experiment having detector efficiency $\eta$, an adversary can use \cref{eqn2} to calculate the maximum correlation value $E$ corresponding to that $\eta$, and then use \cref{ab} to set the values of $a,b$. As long as \cref{ab} is maintained, the single click probability during the test is equal to $\eta$ and the CHSH value is equal to the bound as shown by the thick black dashed line in \cref{fig:result_ce}. For example, for a Bell test done with detector efficiency $\eta = 2(\sqrt{2}-1)$, the local bound is $S'= 2\sqrt{2}$ according to \cref{eqn2}. This can be attained---according to \cref{ab}---if  $a = 12\sqrt{2} - 16 = 0.97$ and $b= 40-28\sqrt{2} = 0.40$, which leads to $|E| = 1/\sqrt{2}$. Similarly, the local bound of $S'$ [\cref{eqn2}] can be achieved for any $2/3 \le \eta \le 1$. For $\eta \le 2/3$, the local bound reaches the algebraic maximum $S' = 4$ [\cref{eqn2}].  Note that this method leads to asymmetric detection efficiency, as Bob's efficiency is always higher than Alice's. However, this can be avoided by reversing the roles of Alice and Bob half of the time. This concludes our local model that can emulate the local bounds given by \cref{eqn2} for every value of $ 2/3 \le \eta \le 1$.

\section*{Conclusion}
\label{conclusion}

Although it is a known fact that a local theory can violate a Bell inequality up to a threshold detection efficiency, it is rarely addressed in the literature how an adversary can actually implement it. In this work, we have shown that the existing detector control method can be exploited to implement a local model that can fake the CHSH Bell inequality [\cref{eqn1}] up to the threshold efficiency. Our model can also simulate the local bound of the Garg-Mermin Bell inequality [\cref{eqn2}] for efficiency over 2/3. Our results point out that whenever Bell violations are used for testing less-conventional theories, implementing device-independent quantum secure communication \cite{acin2007}, certifying randomness \cite{pironio2010} and nonlocality, loophole-free Bell tests \cite{hensen2015,giustina2015,shalm2015} should be performed. We would like to point out that there are Bell inequalities that use non-maximally entangled states with threshold efficiency $\eta_{\text{T}} = 2/3$ \cite{eberhard1993}; however, whether our model is effective against those would be a task for future study. 

\section*{Methods}
\label{satisfy_a3}
\textbf{Strategies for controlling \bm{$a$} and \bm{$b$}}

Here we show that regardless of the value of $\alpha_0$ and $\alpha_1$ an adversary can satisfy the assumption that whenever the Alice's basis matches (does not match) that of the incoming light, a deterministic (random) outcome is produced with probability $a~(b)$. For simplicity, let us assume the case when the adversary sends a light polarized at angle $\alpha_0$ towards Alice (strategies for the other polarizations are similar). Then, with probability $(a-b)$, she sends light polarized at angle $\alpha_0$ which, when measured in the same (different) basis, results in detection (no detection) if intensity is tailored properly (see \cref{tab:third-model-control-ab}). With probability $b/2$, she sends the light at an angle midway between $\alpha_0$ and $\alpha_1~(\alpha_1^\perp)$ at angle $\alpha_0 +\phi_1~(\alpha_0 -\phi_1^\perp)$. Here, $\phi_1 = |\alpha_0 -\alpha_1|/2$, $\phi_1^\perp = |\alpha_0 -\alpha_1^\perp|/2$. As a result, when the basis matches, for both the cases, outcome is $\alpha_0$ while for basis mismatch the outcome is $\alpha_1$ and $\alpha_1^\perp$ with probability $b/2$. The condition for this is $I \sin^2\phi < I_{\text{th}} < I \cos^2(\phi)$ for $\phi \in \{\phi_1,\phi_1^\perp \}$ as shown in \cref{tab:third-model-control-ab}. For the remaining times (with probability $1-a$), the adversary sends vacuum. Overall, from \cref{tab:third-model-control-ab}, it can be seen that when the basis matches that of the incoming light, it results in a deterministic outcome with probability $a$; while when the basis mismatches, it results in a random outcome with probability $b$. This supports the practicality of our assumption. Note that this method leads to asymmetric detection efficiency, as Bob's efficiency is always higher than Alice's. However, this can be avoided by reversing the roles of Alice and Bob half of the time. 

\begin{table*}[b]
	\centering
	\caption{Strategy to practically simulate deterministic (random) outcome with efficiency $a$~($b$). Here, $\phi_0 = |\alpha_0 -\alpha_1|/2$, $\phi_1 = |\alpha_0 -\alpha_1^\perp|/2$, and `x'~represents no detection.}
	\label{tab:third-model-control-ab}
	\renewcommand{\arraystretch}{1.16}
	\begin{tabular}[t]{@{\extracolsep{2ex}}c@{}c@{}c@{}c@{}c@{}c@{}}
		\hline\hline
		\multirow{2}{*}{Probability} & \multirow{2}{*}{Intensity} & \multirow{2}{*}{Polarization} & \multicolumn{2}{c}{Outcome when basis} & \multirow{2}{*}{Required value of $I$} \\
		\cline{4-5}
		&&& matches & mismatches \\
		\hline 
		$a-b$ & $I$    & $\alpha_0$          & $\alpha_0$ & x                & $I \ge I_{\text{th}}$, $I \cos^2(2\phi_0) < I_{\text{th}}$, $I \sin^2(2\phi_0) < I_{\text{th}}$ \\
		$b/2$ & $I$    & $\alpha_0 + \phi_0$ & $\alpha_0$ & $\alpha_1$       & $I \sin^2(\phi_0) < I_{\text{th}} \le I \cos^2(\phi_0)$ \\	          
		$b/2$	& $I$    & $\alpha_0 - \phi_1$ & $\alpha_0$ & $\alpha_1^\perp$ & $I \sin^2(\phi_1) < I_{\text{th}} \le I \cos^2(\phi_1)$ \\
		$1-a$ & vacuum &                     & x          & x                & \\ 
		\hline\hline
	\end{tabular} 
\end{table*}

We have so far assumed that the blinded detector is controllable as a step function: for $I < I_\text{th}$ the click probability is 0, and for $I \ge I_\text{th}$ it is 1. This is of course a simplification \cite{lydersen2010a,gerhardt2011,liu2014,lydersen2010b,wiechers2011,lydersen2011c,sauge2011,jogenfors2015,huang2016,chistiakov2019,gras2020}. Real detectors have noise, which leads to them having two thresholds $I_\text{never} < I_\text{always}$, with click probability 0 for $I \le I_\text{never}$ and 1 for $I \ge I_\text{always}$. In the range $I_\text{never} < I < I_\text{always}$, the click probability gradually increases from 0 to 1. These thresholds depend on the blinding power and regime. Furthermore, no two detector samples are identical, and require tweaking the faked states to achieve perfect or near-perfect control \cite{gerhardt2011,liu2014,jogenfors2015}. Generally, if the ratio $I_\text{always}/I_\text{never}$ can be made sufficiently small, perfect control can be achieved. These issues are device-specific and should be treated at the implementation stage. However, the ability to obtain an arbitrary click probability by adjusting $I$ may allow an alternative method of controlling $a$ and $b$, as we show below.

\begin{figure}
	\centering
	\includegraphics{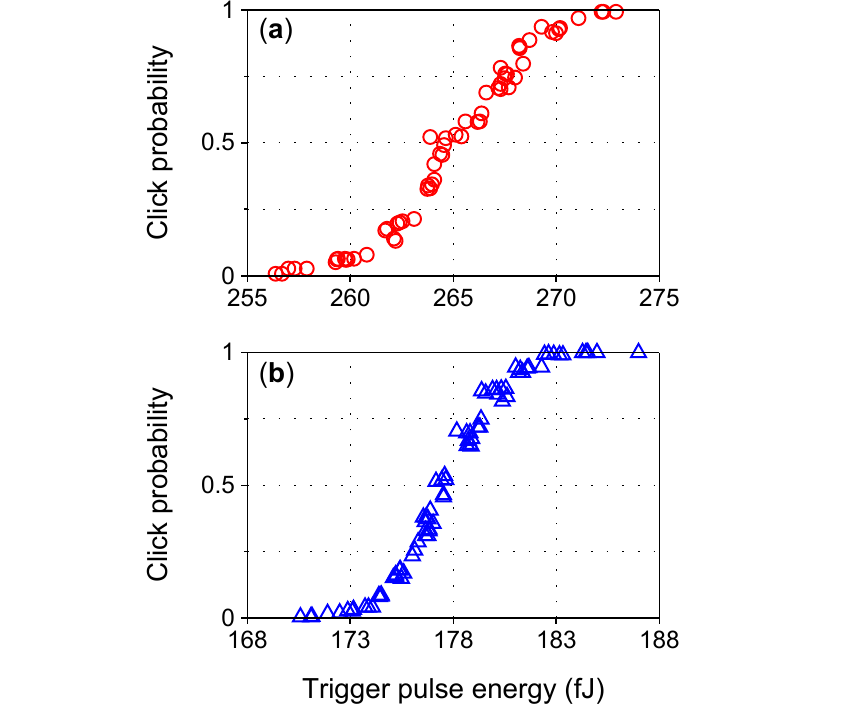}
	\caption{Control characteristics of a detector in commercial quantum key distribution system Clavis2~\cite{idqclavis2specs,huang2016}, responding to a short trigger pulse atop continuous-wave blinding power of (a)~$740~\micro\watt$ and (b)~$367~\micro\watt$. Wavelength of light was $\sim 1.55~\micro\meter$.}
	\label{fig:control-trigger-energy-Clavis2}
\end{figure}

Practical detectors, when blinded, gradually increase their click probability from 0 to 1 in a certain range of trigger intensity $I$ \cite{lydersen2010a,gerhardt2011,liu2014,lydersen2010b,wiechers2011,lydersen2011c,sauge2011,jogenfors2015,huang2016,chistiakov2019,gras2020}. This can be used to obtain probabilistic detections. To illustrate this, we have measured control characteristics of one avalanche photodiode detector in a commercial QKD system Clavis2~\cite{idqclavis2specs,huang2016}. At a particular continuous-wave blinding power, we varied the trigger pulse energy and recorded the corresponding click probability as shown in \cref{fig:control-trigger-energy-Clavis2}. The result shows that it is in principle possible for an adversary to select a value of trigger pulse intensity $I$ (without varying the polarization by $\pm \phi$) that in a matching basis leads to click probability $1$ in one detector, and when halved owing to basis mismatch, leads to a random click in either detector with probability $\sim 0.40$. However, some double clicks (i.e.,\ simultaneous clicks in both detectors) will happen in this strategy. Their handling in a Bell test will need to be considered. 

\def\bibsection{\medskip\begin{center}\rule{0.5\columnwidth}{.8pt}\end{center}\medskip} 


\section*{Acknowledgments}
We thank D.~A.~Graft, G.~Adenier, H.-K.~Lo, and Y.~Zhang for discussions. This work was funded by Industry Canada, CFI, NSERC (programs Discovery and CryptoWorks21), Ontario MRI, US Office of Naval Research, and the Ministry of Education and Science of Russia (program NTI center for quantum communications).

\section*{Author contributions statement}
S.S.\ developed the perfect local model, measured detector characteristics, and wrote the manuscript with input from all authors. N.S.\ developed the improvement to existing model. C.C.W.L.\ contributed to the study of the perfect local model and V.M.\ supervised the study.

\section*{Competing financial interests}
The authors declare no competing financial interests.

\end{document}